\documentclass[12pt]{iopart}
\usepackage{epsfig, graphicx}
\begin{document}

\title[Effect of Ti substitution in La$_{0.67}$Sr$_{0.33}$Mn$_{1-x}$Ti$_x$O$_{3}$] {Effect of  Ti$^{4+}$ substitution on
structural, transport and magnetic properties of La$_{0.67}$Sr$_{0.33}$Mn$_{1-x}$Ti$_x$O$_{3}$.}
\author{Vishwajeet Kulkarni\dag,  K. R. Priolkar\dag\footnote[3]{Author to whom correspondence should be addressed.}, P R Sarode\dag, Rajeev Rawat\ddag,  Alok Banerjee\ddag,  and  S. Emura\P}
\address{\dag\ Department of Physics, Goa University,Taliegao Plateau, Goa 403 206 India}
\address{\ddag\ UGC-DAE Centre for Scientific Research, University Campus, Khandwa Road, Indore 452 017 India}
\address{\P\ Institute of Scientific and Industrial Research, Osaka University, Mihoga-oka 8-1, Ibaraki, Osaka 567-0047, Japan.}

\ead{krp@unigoa.ac.in}
\date{\today}

\begin{abstract}
 La$_{0.67}$Sr$_{0.33}$Mn$_{1-x}$Ti$_x$O$_{3}$  ($0 \le x \le 0.20$)
 polycrystalline materials are prepared by employing lower annealing
 temperature compared to the temperatures reported for the materials
 of the same composition. The transport and magnetic properties of
 these materials are  significantly different from those compounds
 prepared at higher annealing temperature. Samples with $x < 0.10$,
 show metal-insulator transition and those with $x \ge 0.10$ exhibit
 insulating  behavior over the entire temperature range
 investigated. A gradual transition occurs from
 ferromagnetic-metallic state to  ferromagnetic-insulator state with
 increasing Ti substitution. Lattice parameters and bond lengths of
 Mn and its near neighbours however do not change appreciably with
 the dopant content $x$ in these materials. It is shown that
 Ti$^{4+}$ doping in the low temperature annealed samples is
 inhomogeneous resulting in isolated Mn rich regions that are
 connected by a variable range hopping polaron.
\end{abstract}
\submitto{\JPCM}
\maketitle

\section{Introduction}

 Colossal magnetoresistive (CMR) manganites are among the most
 studied materials in condensed matter physics \cite{Colo1,Colo2}.
 Physical properties of manganites are very sensitive to the method
 of preparation, the type of symmetry of a unit cell, A and B site
 cations,  size effects, concentration of the substitute,
 hetrovalent substitution, non stoichiometry etc. Most of the CMR
 properties are explained in terms of Double-exchange(DE) and/or
 superexchange interactions, charge localization via Jahn-Teller
 distortion with polaron formation, phase separation and
 site-disorder. Yet there is no agreement on the correct theoretical
 description of CMR properties, due to the complexities involved.
 Lately, there has been  a greater focus on the effect of the
 synthesis procedures on CMR properties of these materials. It is
 well known that CMR materials exhibit significantly different
 properties when they are prepared with different synthesis
 procedures, starting from the same initial composition.  Various
 defects like cation vacancies, oxygen non-stoichiometry and cation
 site disorder develop in the crystal structure due the synthesis
 procedure and the environment used to prepare these materials.
 However, the effect of synthesis procedure on the microscopic as
 well as macroscopic properties of these materials is quite complex
 and needs to be investigated further to understand the physics
 involved.

Effect of oxygen non-stoichiometry on  properties of CMR samples has
also been studied in great detail \cite{BCTo,NSak,AMai,SGel,JAMv1}.
For example, the defect chemistry of LaMnO$_{3\pm\delta}$ is unique
\cite{JMiz}. For LaMnO$_{3+\delta}$, the structure is GdFeO$_3$-type
for $0 \le \delta < 0.10$ and rhombohedral for $0.10 \le \delta <
0.3$ \cite{Elai}. Although traditionally lanthanum manganite is
considered as an anion excess compound (LaMnO$_{3+\delta}$),
detailed investigations by employing high-resolution electron
microscopy and other cognate techniques have revealed the presence
of metal vacancies (La$_{1-\gamma}$Mn$_{1-\gamma}$O$_3$) instead of
interstitial oxygen ions \cite{MHer,JAMv2}. LaMnO$_3$ perovskites
has been shown to tolerate a considerable portion of vacancies in
the A site (La site) giving rise to compositions of the type
 La$_{1-\delta}$MnO$_3$ with the charge compensated by formation of
 Mn$^{4+}$ \cite{VFer}. The  Mn$^{4+}$ content in lanthanum
 manganites can be varied by altering the firing temperature and
 atmosphere.  Oxygen-rich samples exhibit ferromagnetic-metallic to
 paramagnetic-insulating transition due to the holes doped by cation
 vacancies.  However, in deliberately prepared cation deficient
 material like La$_{1-x}$MnO$_{3+\delta}$, it has been argued that
 charge deficit due to vacancies on La sites would be preferably
 compensated by the formation of oxygen vacancies, rather than by
 oxidation of Mn$^{3+}$ into Mn$^{4+}$ ions. Recently, XRD and EXAFS
 studies \cite{GDez} on vacancy doped La$_{1-x}$MnO$_{3+\delta}$
 have shown that for La/Mn ratio below 0.9, there is phase
 segregation of La$_{0.9}$MnO$_3$ and parasitic Mn$_3$O$_4$ and
 T$_C$ remains almost constant as that of La$_{0.9}$MnO$_3$.

 Strontium doped LaMnO$_3$ series also exhibits both oxygen rich as
well as oxygen deficient non-stoichiometry. Trukhanov et
al\cite{SVTr1,SVTr-} have studied oxygen deficient
La$_{1-x}$Sr$_x$MnO$_{3-{x\over2}}$ compounds wherein ferromagnetic
indirect superexchange is considered to be the cause of  the
magnetism. There are several reports \cite{JAMv1,JAMv2,BDab,ZBuk} on
La$_{1-x}$Sr$_x$MnO$_{3+{\delta}}$ compounds, wherein cation
vacancies in oxygen rich materials are considered to be equally
distributed on A and B sites, given by $ v = {\delta \over
(3+\delta)}$ whereas Tofield and Scott \cite{BCTo} and Mitchell {\em
et al} \cite{JFMi} have shown that the A-site vacancies are
predominant. Mizusaki {\em et al} have discussed  various models for
cation deficiency in La$_{1-x}$A$_x$MnO$_3$ \cite{Juni} such as
A-site cation substituting B-site and vice-versa. However, there
appears to be very little literature about B-site vacancies
\cite{JAMR,JToe} in these materials. Interestingly, Nakamura
\cite{Keik} has shown that a fraction of Mn ions enter into A-site
cation vacancies in La$_{1-\Delta}$MnO$_{3+\delta}$ and affects
properties of the material substantially.

 In spite of the extensive work carried out on these materials,
there is no clear understanding and consensus on the microscopic
picture of the disorder caused by vacancies or dopant atoms in these
materials. Moreover, the effect of  dopant ions replacing
hetrovalent ions  has not been studied so far. Such a disorder can
have tremendous effect on the properties of these materials and is
expected to give some interesting results. Ti doped CMR manganites
offer such an opportunity.  Ionic radius of Ti$^{4+}$ ion is known
to be in between those of Mn$^{4+}$ and Mn$^{3+}$ ionic radii and
there exists a distinct possibility that a fraction of Ti$^{4+}$
ions substitute for Mn$^{3+}$ ions leading to oxygen non
stoichiometry (cation deficiency) or such inhomogeneities. The
available literature on Ti substitution in manganites indicates that
Ti$^{4+}$ ions substitute the isovalent Mn$^{4+}$ ions in these
materials \cite{sah,hu,troy,kal,liu,kim,uly1,uly2,alva,nam}.
Recently however, it has been shown that at high doping levels
Ti$^{4+}$ ions occupy Mn$^{3+}$ sites in addition to Mn$^{4+}$ sites
\cite{xx}. Effect of lower temperature annealing has also been
studied recently \cite{krp}. It has been shown that by suitably
modifying the preparation procedure, Ti$^{4+}$ ions can substitute
Mn ions randomly in La$_{0.67}$Sr$_{0.33}$Mn$_{1-x}$Ti$_x$O$_3$
forming inhomogeneous short range ordered ferromagnetic clusters.
Here we report a detailed investigation of structural, magnetic,
transport and spectroscopic properties of lower temperature annealed
La$_{0.67}$Sr$_{0.33}$Mn$_{1-x}$Ti$_x$O$_3$ ($0 \le x \le 0.2$)
using techniques like XRD, resistivity, AC and DC susceptibilities,
XPS, Mn and La K-edge EXAFS and IR spectroscopy. Since annealing
temperature of these materials is substantially lower than that
reported in \cite{kal,kim}, we term these as low-temperature
annealed materials. For comparison, $x = 0.10$ sample has been
prepared at the same annealing temperature as in \cite{kal} and is
referred to as high-temperature annealed  sample.

\section{Experimental}
 Polycrystalline samples of
La$_{0.67}$Sr$_{0.33}$Mn$_{1-x}$Ti$_x$O$_3$ with $x$ = 0.0, 0.03,
0.05, 0.10 and 0.20  were synthesized by conventional solid-state
reaction method. The powders of La$_2$O$_3$, SrCO$_3$, freshly
prepared MnCO$_3$ and TiO$_2$ with proper stoichiometry were  mixed,
ground manually using agate mortar and pestle and fired  in  air at
1000$^\circ$C for about 15 hours, reground to calcine at
1100$^\circ$C for 20 hours. Finally they were pressed into pellets
and sintered in air at 1200$^\circ$C for 20 hours.  Grinding period
was about half hour every time. The $x = 0.10$ sample  was prepared
also by firing the ingredients in stoichiometric proportions at
1000$^\circ$C, 1200$^\circ$C and annealing at 1450$^\circ$C which
closely match with those reported in \cite{kal,kim}. In this case
however the intermediate grinding was done manually with agate
mortar and pestle for about 3 hours.

 The samples were  characterized  by  X-ray  powder  diffraction
recorded on Siemens D5000 diffractometer at  room temperature
between $2\theta$  = 10$^\circ$ to 70$^\circ$ with a step
0.05$^\circ$. For AC susceptibility measurements, the samples were
cooled under zero magnetic field and measurements were carried  out
in the warming run under the applied field of  0.936 Oe and AC
frequency 133.33 Hz from liquid nitrogen temperature to 310 K on an
in-house AC Susceptometer \cite{bajp}. For the samples with x = 0.00
and x = 0.03,  the susceptibility measurements were carried out up
to 365 K and 330 K respectively. Standard four-probe technique was
used to measure resistivity of the samples in the temperature
interval 300 K to 80 K. Magnetoresistance (MR) measurements were
performed down to 30 K using the standard four-probe geometry in
transverse magnetic fields up to 5 T using OXFORD Spectralab 10 T
superconducting magnet. Infrared measurements were performed on
Shimadzu FTIR-8900 spectrophotometer at room temperature in
transmission mode in the range of 350 cm$^{-1}$ to 1000 cm$^{-1}$.
The samples were mixed with KBr in the ratio 1:100 by weight and
pressed into pellets for IR measurements. XPS study was carried out
in an ESCA-3 Mark II spectrometer (VG Scientific Ltd., England)
employing Al K$_\alpha$ radiation (1486.6 eV) at a pass energy of 50
eV. The powder samples were made into pellets of 8 mm diameter and
placed into an UHV chamber housing the analyser at 10$^{-9}$ Torr.
Before the measurements were carried out, the samples were kept at
the preparation chamber for 5 hours for desorption of gases. Binding
energies were measured with a precision of $\pm 0.1$eV. The charging
effect was taken care of with respect to the C(1s) peak of
adventitious carbon at 285 eV.   XAFS spectra at La and Mn K-edges
were recorded using BL01B1 XAFS beam-line at SPring-8.  Si(111)
crystal served as the monochromator for Mn K-edge EXAFS and Si(311)
for La K-edge EXAFS.  For Mn K-edge measurements, fine powder of the
sample was brushed onto scotch tape.  A number of layers of tape
were stacked to obtain  total absorption lengths $\mu x$ $\approx$
2.5 ( where $x$ is the thickness of the sample ). For La K-edge
measurements, the absorbers were made by pressing the samples into
pellets of 10 mm diameter with boron nitride as binder. The
thickness(x) of the absorber was adjusted such that $\mu x$ $\ge$ 1.
The pre-edge absorption was removed by fitting the data to a linear
background and a simple cubic spline was used to simulate the
embedded-atom absorption, $\mu_0$, above the edge. The XAFS
oscillations $\chi$ were obtained as a function of  photoelectron
wave vector $k = \sqrt{2m(E - E_0)/h^2}$. E$_0$ was estimated from
the first inflection point of the main edge. XAFS oscillations,
$\chi(k)=(\mu-\mu_0)/\Delta\mu$ (k - photoelectron wavenumber,
$\mu_0$ - atomic absorption coefficient and $\Delta\mu$ - edge
jump), were extracted following standard procedures. FEFFIT program
\cite{MNew} was used to fit Fourier transformed (FT)  k$\chi(k)$
data in  $r$ space to the theoretical spectra calculated using
FEFF6.01 \cite{JJRe}. The Mn and La K-edge EXAFS data were analysed
in tandem, to improve the reliability of fits.

\section{Results}
\subsection{X-ray Diffraction}
\begin{figure}
\centering
\epsfig{file=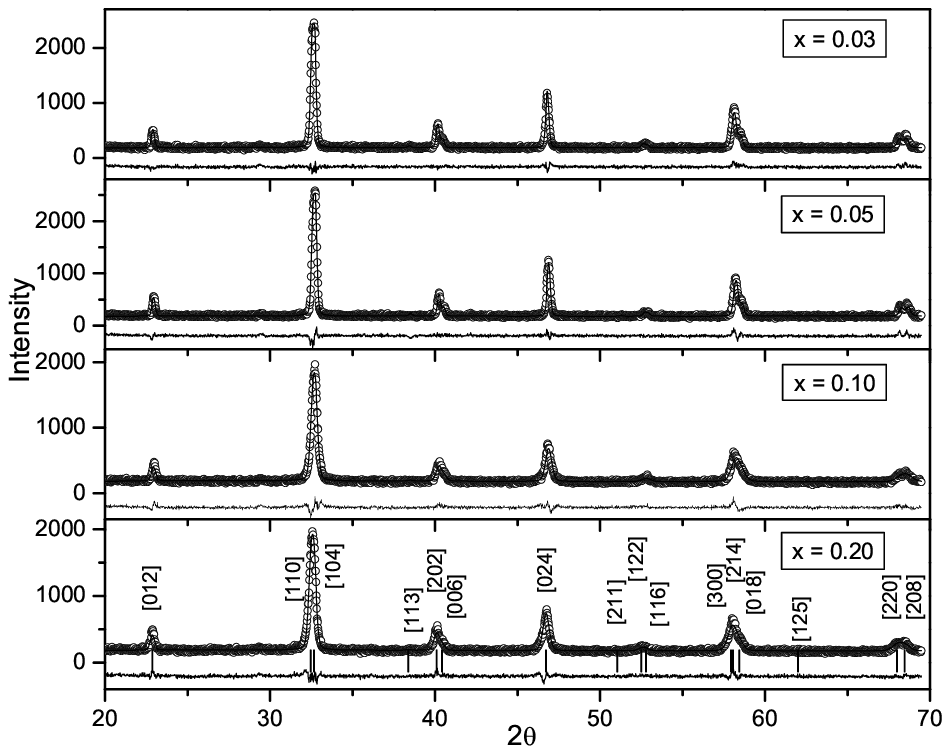, width=18cm, height=18cm}
\caption{X-ray diffraction patterns (shown with circles)  for low-temperature annealed La$_{0.67}$Sr$_{0.33}$Mn$_{1-x}$Ti$_x$O$_{3+\delta}$ ($0 \le x \le 0.20$)samples and their Rietveld fits (shown with line). Vertical lines in the bottom panel indicate the Bragg reflection positions.}
\label{fig:xrd}
\end{figure}

The X-ray diffraction patterns of
La$_{0.67}$Sr$_{0.33}$Mn$_{1-x}$Ti$_x$O$_{3+\delta}$  ($0 \le x \le
0.20$) and the fits obtained from Rietveld refinement of the X-ray
diffraction data are presented in Fig.\ref{fig:xrd}. All the samples
crystallize in a single phase rhombohedral structure (space group
R\={3}C). Lattice parameters calculated for hexagonal setting of the
space group are found to be $a$ = 5.486\AA ~ and  $c$ = 13.345 \AA~
for the pristine sample [Table \ref{magn}]. These values closely
match with those reported earlier \cite{kal}. A monotonous increase
in lattice parameters has been observed by Kallel et al \cite{kal}
and Kim et al \cite{kim} for increasing doping content $x$ in their
Ti doped La$_{0.7}$Sr$_{0.3}$MnO$_3$ samples. They have ascribed the
increase in lattice parameters to  larger Ti$^{4+}$ ions
substituting smaller Mn$^{4+}$ ions. In our samples, we observe
non-systematic and relatively insignificant changes in the lattice
parameters with the doping content $x$. Recently, Zhu et al
\cite{xx} have reported that in thin films of
La$_{0.67}$Sr$_{0.33}$Mn$_{1-x}$Ti$_x$O$_{3+\delta}$, the lattice
parameters initially increase with doping content $x$ for $x \le
0.3$ and decrease for further doping. The decrease in the lattice
parameters is said to be due to smaller Ti$^{4+}$ ions substituting
larger Mn$^{3+}$ ions. The  nearly constant lattice parameters
observed in our  samples, could therefore be possible if  Ti$^{4+}$
ions substitute both larger Mn$^{3+}$ and smaller Mn$^{4+}$ ions at
random.

\begin{table}
 \caption{\label{magn} Doping content, excess oxygen,
lattice-parameters $a$ and $c$, Susceptibility, calculated spin-only
susceptibility, and T$_{C}$ for low-temperature annealed
La$_{0.67}$Sr$_{0.33}$Mn$_{1-x}$Ti$_x$O$_{3+\delta}$. * sign implies
VRH supported ferromagnetic transition around room temperature.}
\begin{indented}
\item \begin{tabular}{@{}ccccccc}
\br
x & $\delta$ & $a$(\AA) & $c$(\AA) & $\chi$$_{exp}$ &$\chi$$_{spin-only}$ & T$_{C}$(K)\\
& & & &($\mu_B$/f.u.-Oe)&($\mu_B$/f.u.-Oe)&\\
\mr
00&0.00&5.49(1)&13.35(1)&3.53&3.67&360\\
0.03&0.01&5.49(1)&13.36(1)&1.79&3.58&325\\
0.05&0.02&5.49(1)&13.34(1)&2.55&3.52&305\\
0.10&0.03&5.49(1)&13.34(1)&0.75&3.37&$\sim$300*\\
0.20&0.05&5.50(1)&13.36(1)&0.76&3.07&$\sim$300*\\
\br
\end{tabular}
\end{indented}
\end{table}

\subsection{Infra-red Absorption Studies}
\begin{figure}
 \centering
\epsfig{file=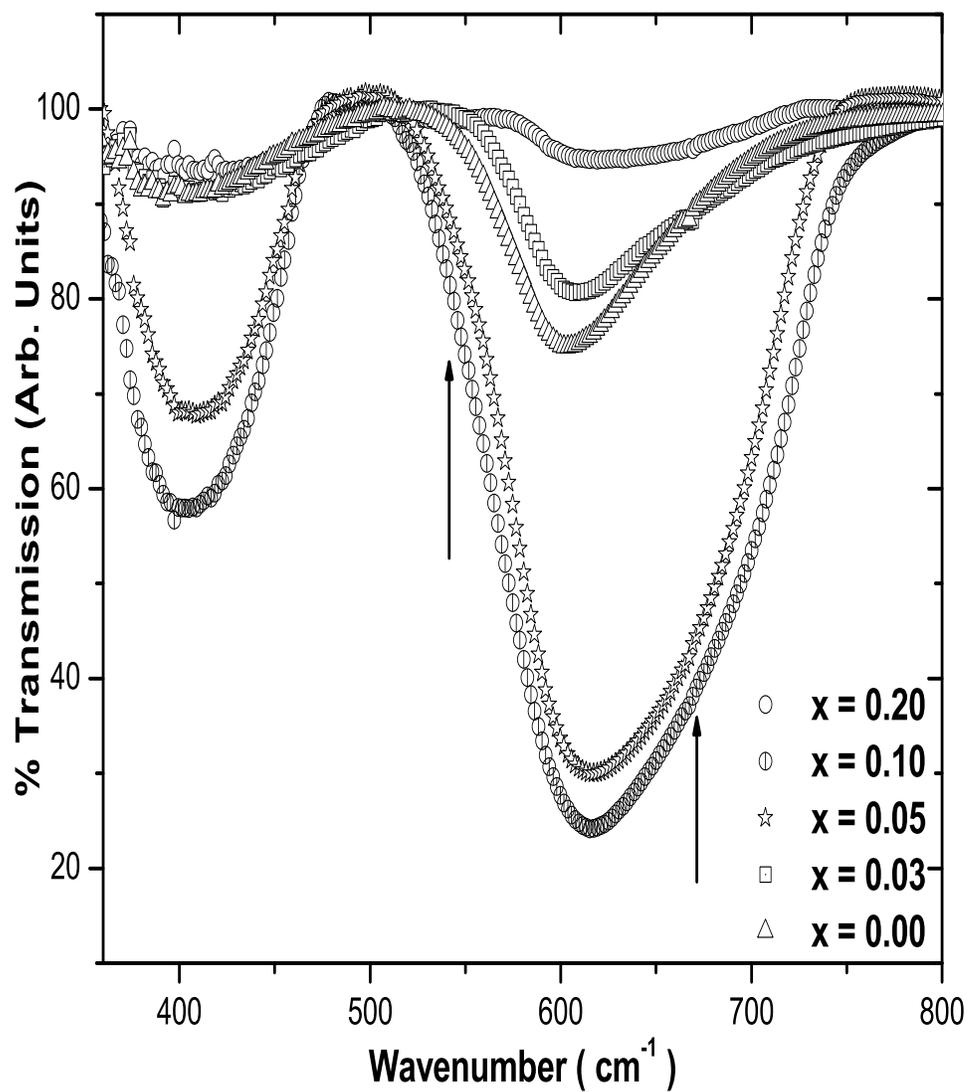, width=15cm, height=18cm}
\caption{IR transmission spectra of low-temperature annealed  La$_{0.67}$Sr$_{0.33}$Mn$_{1-x}$Ti$_x$O$_{3}$  ($0 \le x \le
  0.20$).  Arrows indicate Ti-O stretching modes.}
 \label{fig:ir}
\end{figure}

To investigate the valence of Ti ions, IR transmission spectra of
the samples were recorded [Fig.\ \ref{fig:ir}]. For the undoped
sample, two broad absorption peaks are clearly seen between 350-500
cm$^{-1}$ and 500-750 cm$^{-1}$, respectively.  The first peak
between 350-500 cm$^{-1}$ is attributed to Mn-O-Mn bending mode  and
the second to Mn-O stretching mode of vibration \cite{ZVPo,GDeM}.
The widening of the stretching peak with doping content $x$ may be
induced due to the Mn-O and Ti-O stretching vibrations in Mn-O-Ti
structure. Two additional shoulders at 670 cm$^{-1}$ and 540
cm$^{-1}$ appear in heavily doped Ti samples. These  frequencies
match closely with TiO$_{2}$ absorption peaks \cite{Xian} indicating
that Ti is primarily in tetravalent state. Recently temperature
dependent IR absorption studies have been reported on some of these
compounds \cite{krp}.

\subsection{Magnetic susceptibility}
\begin{figure}
 \centering
 \epsfig{file=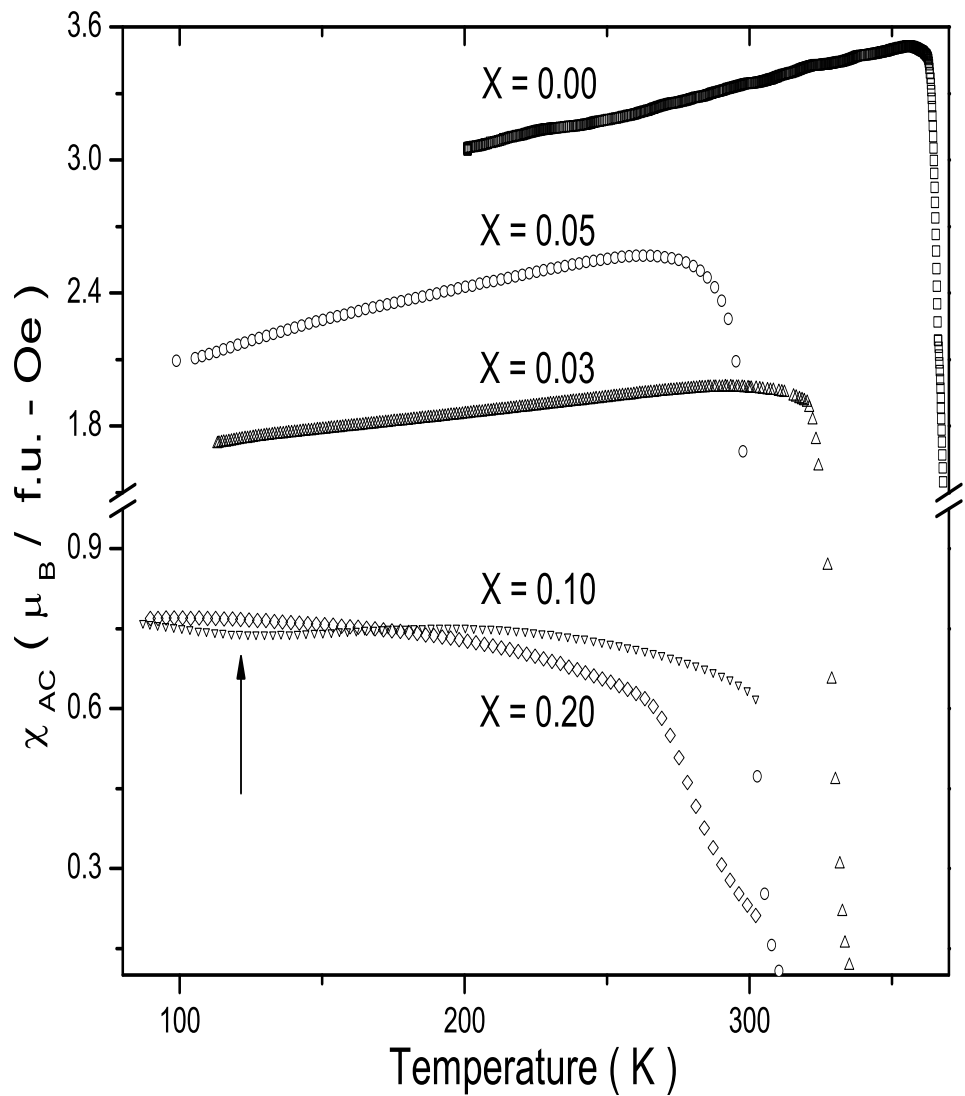, width=15cm, height=18cm}
  \caption{AC susceptibility-temperature plots of low-temperature annealed La$_{0.67}$Sr$_{0.33}$Mn$_{1-x}$Ti$_x$O$_{3+\delta}$ ($0 \le x \le 0.20$). Arrow indicates the second ferromagnetic transition in $x = 0.10$ sample. }
  \label{fig:sus}
\end{figure}

 Plots of AC magnetic susceptibility as a function of temperature
for La$_{0.67}$Sr$_{0.33}$Mn$_{1-x}$Ti$_x$O$_{3}$  ($0 \le x \le
0.20$) samples are presented in Fig.\ \ref{fig:sus}. For  $0.0 \le x
\le 0.05$, a sharp rise in AC susceptibility is seen indicating
paramagnetic to  ferromagnetic  transition. The ferromagnetic
ordering temperature ($T_C$) decreases with increasing doping
content.  $T_C$ obtained for the parent compound is close to 360 K
which is as reported earlier \cite{RMah2} and it decreases to 325 K
and 305 K for $x$ = 0.03 and 0.05 respectively [Table\ \ref{magn}].
However, AC susceptibility of $x = 0.05$ sample is greater than that
of $x = 0.03$ sample. Such anomalies have been reported
earlier\cite{JBla,ASun}. Blasco et al\cite{JBla} have attributed it
to the magnetic inhomogeneity in their samples. In our case too, the
anomaly may perhaps be due to the non-uniform distribution of
magnetic Mn$^{3+/4+}$ ions in the sample. It is also possible that
the Mn$^{3+}$ content in $x = 0.03$ sample is smaller than that in
$x = 0.05$ sample. Such a possibility does exist and can be seen in
the error estimates of Mn$^{3+}$ and Mn$^{4+}$ contents in the two
samples [Table\ \ref{xps}]. For $x$ = 0.10,  AC susceptibility
exhibits two distinct features close to 300 K and 120 K
respectively. This sample has a ferromagnetic like transition just
above 300 K but the magnitude of the susceptibility per formula unit
(f.u.) is very much less than the calculated spin only moment value
for a fully ordered sample [Table\ \ref{magn}].  Susceptibility
again shows a tendency to increase around 120 K. Such a behavior has
been reported in La$_{1.4}$Sr$_{1.6}$Mn$_{2-y}$Ti$_{y}$O$_{7}$
\cite{Hong}. For $x$ = 0.2, a similar weak and broad magnetic
transition is observed at about 300 K with slope of the ac
susceptibility-temperature curve being negative over the entire
temperature range. It may be mentioned here that the calculated
moment values shown in Table \ref{magn} are with an assumption that
Ti$^{4+}$ ions substitute only Mn$^{4+}$ ions.

\begin{figure}
 \centering
\epsfig{file=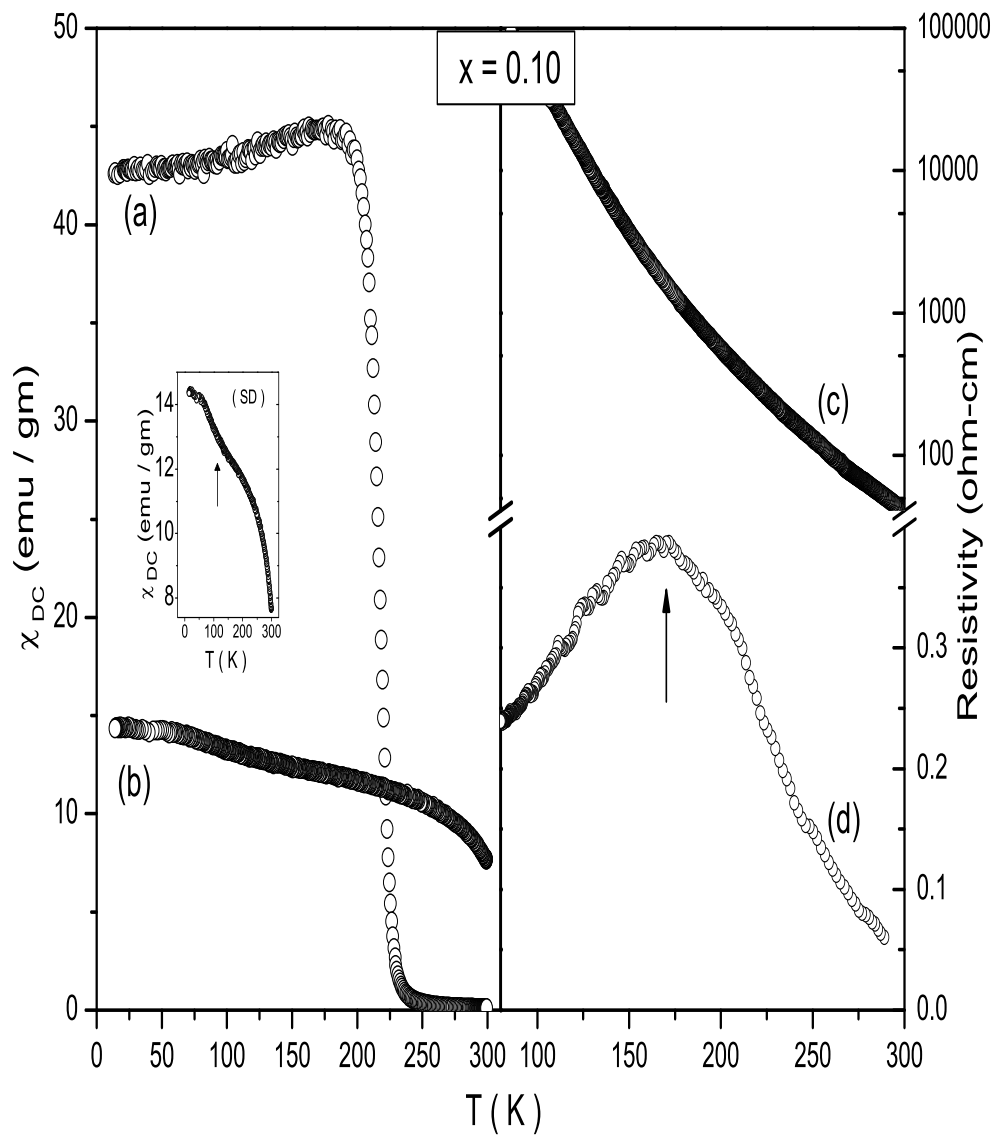, width=15cm, height=18cm}
\caption{Temperature dependence of DC susceptibility for  La$_{0.67}$Sr$_{0.33}$Mn$_{0.90}$Ti$_{0.10}$O$_{3+\delta}$ (a)  high-temperature annealed, (b) low-temperature annealed samples and resistivity of (c) low-temperature annealed and  (d) low-temperature annealed samples. Second ferromagnetic transition in the low-temperature annealed sample is indicated by an arrow in the inset.}
 \label{fig:susres10}
\end{figure}

 To  investigate the  magnetic properties further, DC susceptibility
measurements were carried out in the temperature range of 300 K to
10 K in an applied field of 100 Oe on $0.00 \le x \le 0.20$ samples,
using a Faraday balance. Similar measurements were also carried out
on the high-temperature annealed $x = 0.10$ sample. Susceptibility
of the high-temperature annealed  sample with $x = 0.10$ shows
single ferromagnetic transition at temperature 220 K [Fig.\
\ref{fig:susres10}(a)] whereas corresponding low-temperature
annealed sample shows two transition like-features at 300 K and 120
K respectively [Fig.\ \ref{fig:susres10}(b)].  Susceptibility at
saturation is substantially more for high-temperature annealed
sample compared to that of the corresponding low-temperature
annealed sample. The weak susceptibility observed around 300 K in
the low-temperature annealed sample is not visible in the
high-temperature annealed sample.

 In other Ti doped La$_{1-x}$B$_{x}$MnO$_{3}$ (B = Ca, Pb, Sr)
compounds, in which Ti$^{4+}$ ions is reported to substitute only
Mn$^{4+}$ ions, $T_C$ continuously decreases as a function of dopant
concentration $x$. This is due to the weakening of double exchange
interaction between Mn$^{3+}$-O-Mn$^{4+}$ chains. In the present
case, the $T_C$ initially decreases up to $x = 0.05$ and then
remains at about 300 K for the further doping $0.10 \ge x \ge 0.20$.

 Excess oxygen concentration or equivalently cation
vacancies\cite{SHeb} in the samples and  non-magnetic Ti ions
isolate the regions of DE pairs from one another and results in
lower susceptibility of the samples. Yet, higher $T_C$ for $x \ge
0.10$ samples observed in the samples $x \ge 0.10$ could be possible
if such isolated DE pairs are ferromagnetically linked with one
another via a hopping polaron. Polaronic conduction could probably
answer the higher T$_{C}$ for $x \ge 0.10$ samples. In order to
ascertain the exact mechanism responsible for higher T$_{C}$ for $x
\ge 0.10$ samples,  transport properties of this series have been
studied.

\subsection{Resistivity}
\begin{figure}
\centering
\epsfig{file=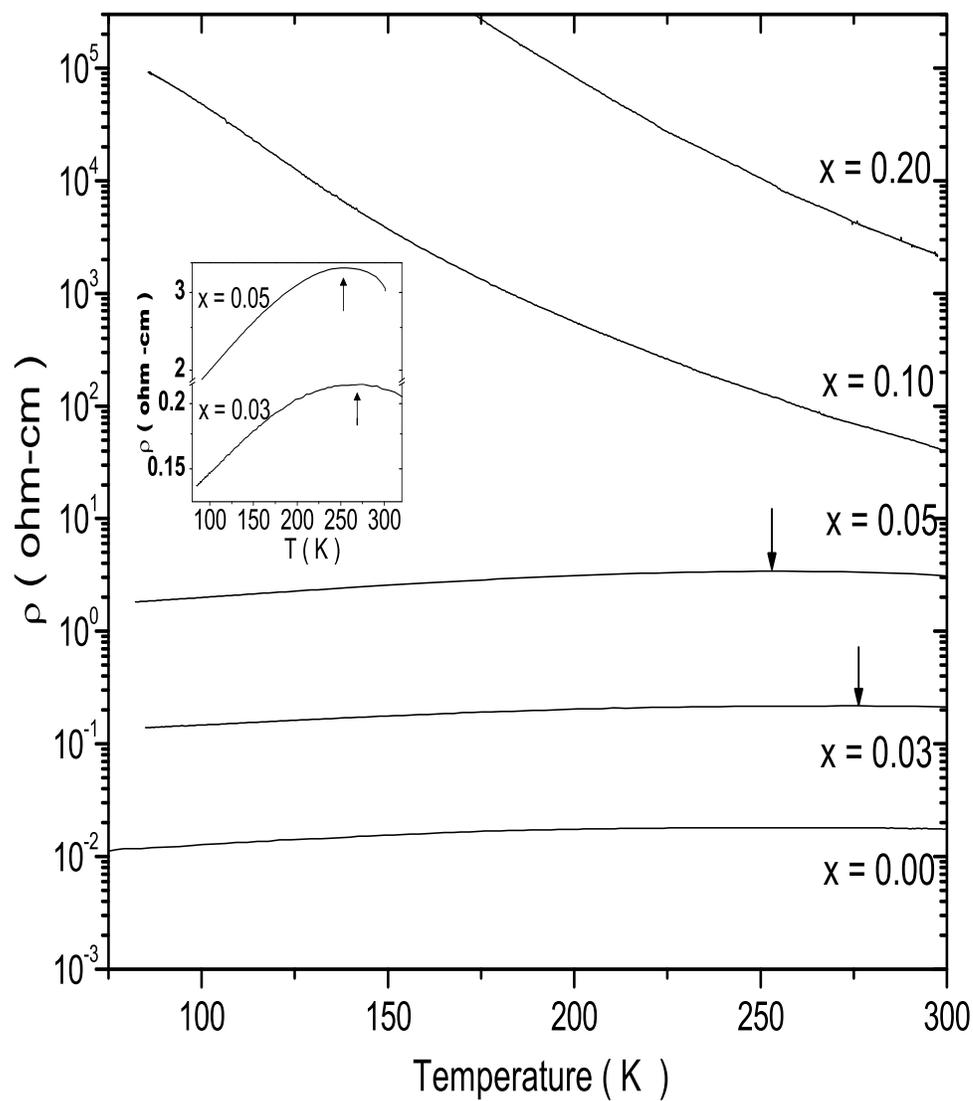, width=15cm, height=18cm}
\caption{Resistivity versus temperature plots of low-temperature annealed La$_{0.67}$Sr$_{0.33}$Mn$_{1-x}$Ti$_x$O$_{3+\delta}$ ($0 \le x \le 0.20$). Arrows in the inset show the  metal-insulator transition in $x$ = 0.03 and 0.05 samples.  }
 \label{fig:res}
\end{figure}

 The resistivity as a function of temperature is plotted in Fig.\
\ref{fig:res} for the Ti doped samples along with the parent
compound  La$_{0.67}$Sr$_{0.33}$MnO$_{3}$. The parent compound shows
a metallic behavior in the entire temperature range studied which
augurs well with its metal-insulator (M-I) transition temperature of
about 330 K \cite{RMah2}. At $x$ = 0.03 and 0.05,  M-I transition
takes place at 270 K and 260 K respectively. For $x \ge 0.10$,
low-temperature annealed sample exhibits insulating behavior
throughout the temperature range studied  while the
high-temperature annealed  sample ($x = 0.10$) shows the MI
transition at 180 K   [Fig.\ \ref{fig:susres10}(c)]. Increase in
resistivity with doping content $x$ in the low-temperature annealed
samples is mainly due to the weakening of the  Mn$^{3+}$-O-Mn$^{4+}$
DE bonds  by interspersed Ti ions. At $x \ge 0.10$, the
ferromagnetic double-exchange energy reduces to the extent that it
does not become comparable to the thermal energy  at any temperature
down to 80 K so as to bring about the M-I transition. It is to be
noted that corresponding high-temperature annealed sample exhibits
M-I transition.

\begin{figure}
 \centering
\epsfig{file=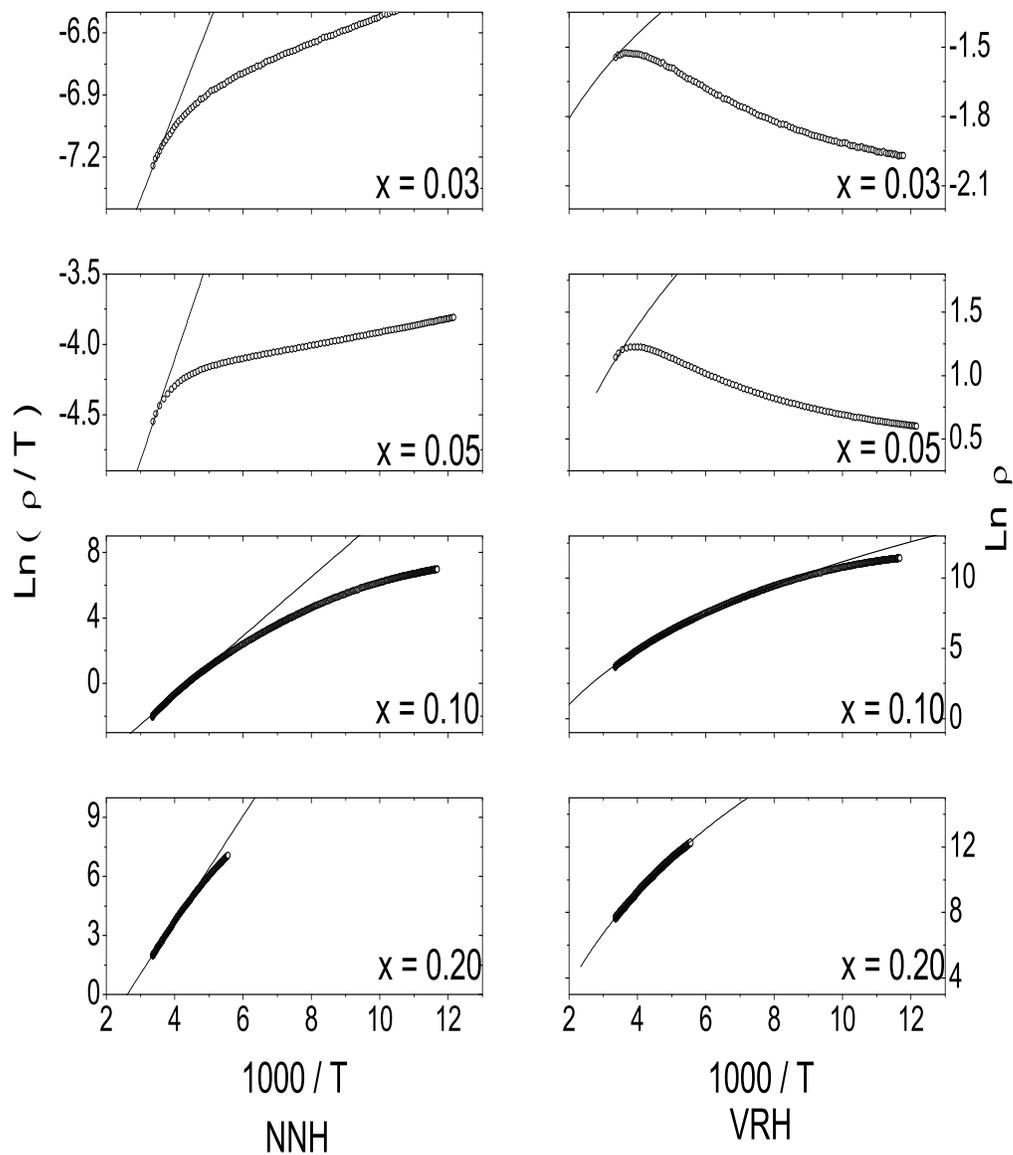, width=15cm, height=18cm}
\caption{Nearest-neighbour and Variable Range Hopping fits shown with
  ln$\rho$ and ln ($\rho$/T) as function of (1000 / T ) for low-temperature annealed La$_{0.67}$Sr$_{0.33}$Mn$_{1-x}$Ti$_x$O$_{3+\delta}$ ($0.03 \le x \le 0.20$)}
 \label{fig:vrh}
\end{figure}

 To study the transport mechanism of the charge carriers, generally
three models are discussed for semiconductor like behavior; the
band-gap model \cite{RMKu}, a nearest neighbour hopping (NNH) model
\cite{GJSn} for the transport of small polarons and Variable Range
Hopping(VRH) model \cite{MVir}. In the band gap model, resistivity
of the sample is given by Arrhenius law $\rho = \rho_{0}
\exp(E{_\rho}/k{_B}T)$. This equation could not be fitted to the
resistivity curves of our low temperature annealed  samples
indicating that the activated transport across the band gap does not
occur in these samples in the temperature range studied. For the NNH
model, the resistivity is given by $\rho = AT \exp(W_{P}/k{_B}T)$
with $ W_{P} = E_{P}/(2 - t)$. $E_{P}$ denotes the polaron-formation
energy and $t$ the electronic-transfer integral. In Mott's VRH
model, the resistivity is expressed in the form of $\rho =
\rho_{inf} \exp(T{_0}/T)^{1/4}$ \cite{Hong}. Figure\ \ref{fig:vrh}
shows that  up to $x = 0.05$,  the transport of the charge carriers
in the insulating region can be explained by  both, NNH and VRH. For
$x = 0.20$, resistivity can be fitted only by VRH equation and for
$x = 0.10$, resistivity deviates from VRH behavior below 130 K. For
$x = 0.20$ resistivity could not be measured below 180 K due to its
very high value. The details of the fitting ranges are presented in
Table\ \ref{nnhvrh}. Further, it may be noted that for $x = 0.10$
sample, the deviation of resistivity from VRH curve at 130 K matches
closely with the second transition observed  in susceptibility plots
in Figs.\ \ref{fig:sus} and \ref{fig:susres10}(b). The weak
ferromagnetic transition around 300 K could then be due to the
interaction between Mn ions via a hopping polaron.  Below 130 K the
hopping energy weakens and Mn rich ferromagnetic regions get
isolated.  These isolated ferromagnetic regions then order  leading
to the second ferromagnetic transition. In sample with $x = 0.20$,
perhaps VRH mechanism extends to a still lower temperature and the
second transition is therefore not distinctly observed.

 It is to be noted that the charge-transport in the high-temperature
annealed $x = 0.10$ sample clearly shows activated behavior across a
band-gap. An exponential curve could be fitted to the
resistivity-temperature plot of this sample between 228 K to room
temperature. Thus the absence of M-I transition in the low
temperature annealed $x$ = 0.1 sample can be understood to be due to
inhomogeneous substitution of Ti$^{4+}$ ions. Whereas in the high
temperature annealed samples, the decrease in the metal to insulator
transition temperature with increasing Ti doping level can be
ascribed to the replacement of some of the Mn$^{3+}$-O-Mn$^{4+}$
bonds by the Mn$^{3+}$-O-Ti$^{4+}$ bonds. This could be either due
to  grain boundary effect or the disorder in Ti doping at lower
annealing temperatures. If grain boundaries are the cause of higher
resistance, then the volume fraction sensitive properties like
magnetization should not be affected \cite{hwang1}. However, it can
be clearly seen that the behavior of magnetization for $x$ = 0.1 of
both high temperature annealed and low temperature annealed samples
is different. Hence grain boundary effect cab be ruled out. A
disorder in Ti substitution could create isolated
Mn$^{3+}$-O-Mn$^{4+}$ pairs connected through a polaron with a
variable hopping range. This fact highlights the importance of VRH
in transport and magnetic properties, and especially, high T$_{C}$,
in $x \ge 0.10$  low-temperature annealed samples.

\begin{table}
 \caption{\label{nnhvrh} Doping content and lower  temperature limit
down to which NNH or VRH fit is possible for low-temperature
annealed La$_{0.67}$Sr$_{0.33}$Mn$_{1-x}$Ti$_x$O$_{3+\delta}$,
higher limit being 300 K.   *Resistivity could not be measured below
180 K due to the high resistivity of the sample.}
\begin{indented}
 \item \begin{tabular}{@{}ccc}
\br
x & NNH & VRH\\
\mr
0.03&276 K&291 K\\
0.05&282 K&280 K\\
0.10&217 K&130 K\\
0.20&236 K&Entire range*\\
\br
\end{tabular}
\end{indented}
\end{table}

\subsection{Magnetoresistance (MR)}
\begin{figure}
 \centering
  \epsfig{file=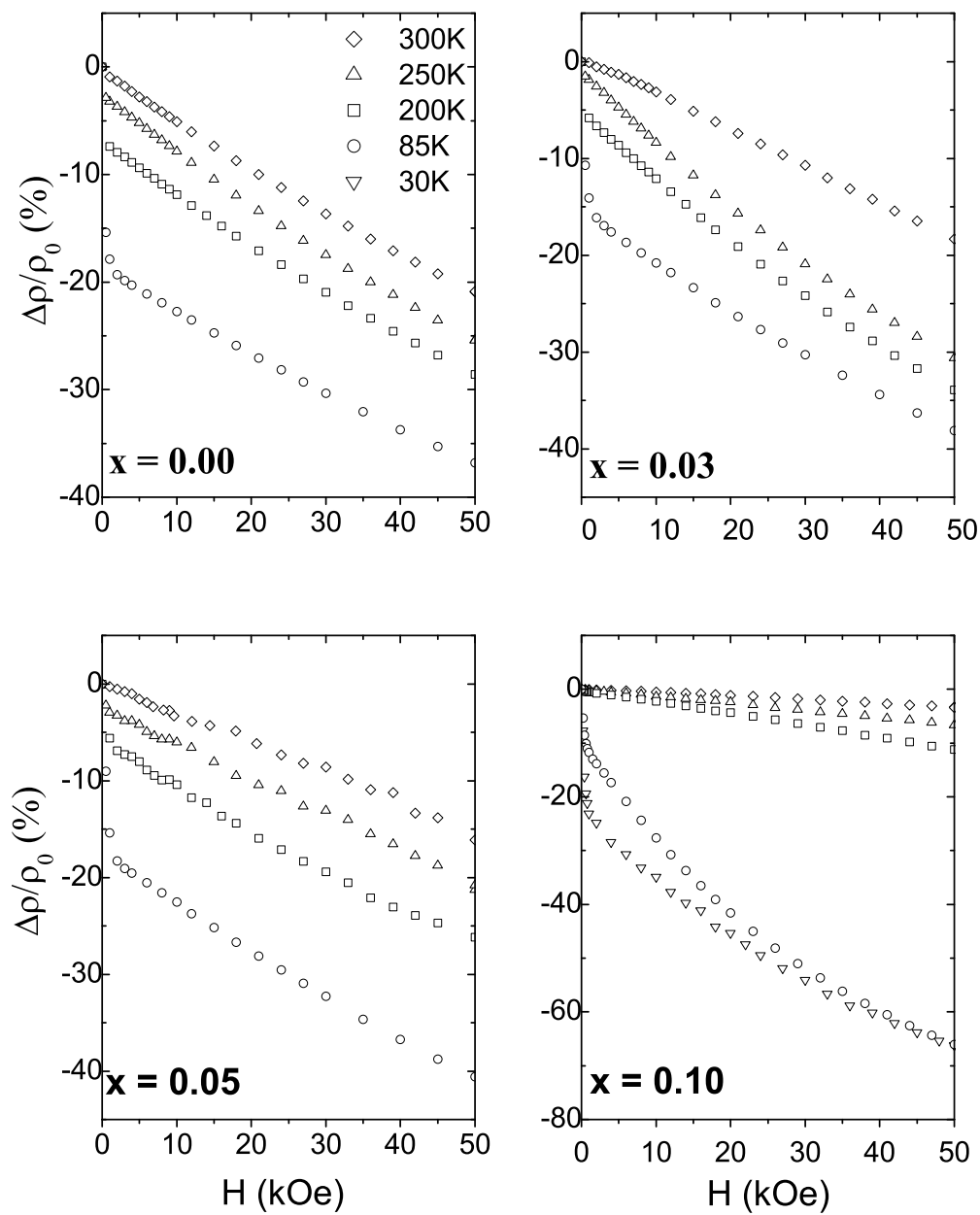, width=15cm, height=18cm}
\caption{Magnetoresistance versus applied magnetic field plots of low-temperature annealed La$_{0.67}$Sr$_{0.33}$Mn$_{1-x}$Ti$_x$O$_{3+\delta}$ ($0 \le x \le 0.10$)}
\label{fig:mr}
\end{figure}

\begin{figure}
 \centering
 \epsfig{file=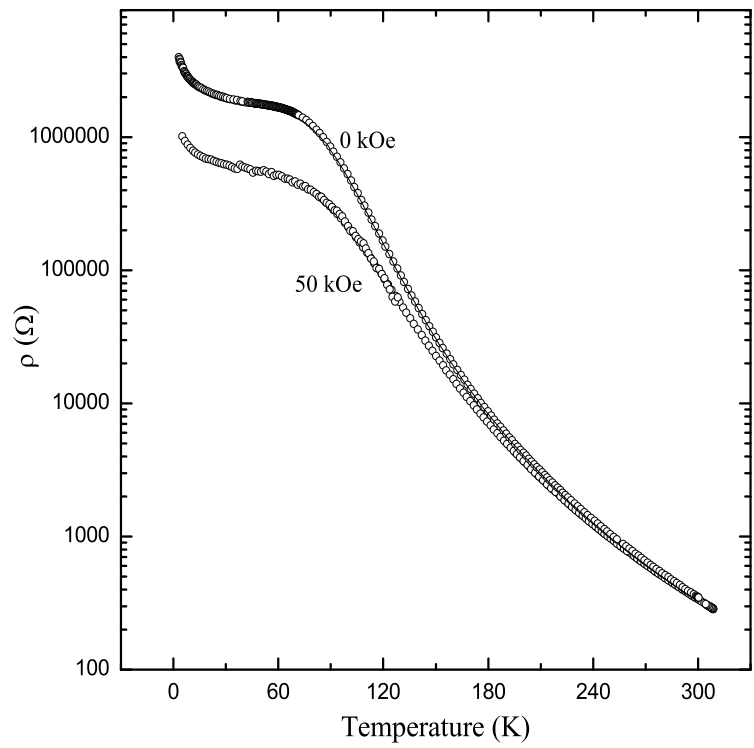, width=12cm, height=12cm}
\caption{Resistivity under zero-field and 5T field for low-temperature annealed x = 0.10 sample. }
\label{fig:RunderH}
\end{figure}

 Isothermal MR is negative for  all the low-temperature annealed
samples ($0.00 \ge x \ge 0.20$) [ Fig.\ \ref{fig:mr}]. For $x =
0.00$, MR shows H$^{2}$ dependence in all the isothermals which is
consistent with the ferromagnetic order of the sample.
Magnetoresistance of La$_{0.7}$Sr$_{0.3}$MnO$_{3}$ has been
reported\cite{RMah2} to be 45\% percent at room temperature  under
the field of 6 T. Magnetoresistance observed for the parent sample
La$_{0.67}$Sr$_{0.33}$MnO$_{3}$ in our series is 38\% at room
temperature and under the field of 5 T. The extrapolation of the
isothermal to 6 T shows that the MR observed matches closely with
that reported in  \cite{RMah2}. Isothermal MR for samples with  $x$
= 0.00, 0.03 and 0.05 increases gradually  at all the temperatures
at  which MR is recorded.  However for $x = 0.10$, the isothermal MR
is smaller than those of $x$ = 0.00, 0.03 and 0.05 up to 210 K and
increases substantially between temperatures 210 K and 85 K. The
sudden increase in MR points towards a ferromagnetic transition
within this temperature range. It is clear from the susceptibility
plots [Fig.\ \ref{fig:sus} and \ref{fig:susres10}(b)] that this
sample has a second magnetic transition at about 120 K. The plot of
resistivity in presence of magnetic field (H = 5 T) also shows a
deviation at about the same temperature from the zero field
resistivity curve  [Fig.\ \ref{fig:RunderH}].

 MR is shown to be much higher at ferromagnetic transition
temperature in the case of La$_{0.815}$Sr$_{0.185}$MnO$_3$
\cite{ZBuk} and cation deficient LaMnO$_3$ \cite{SHeb}. In this case
however, MR for the low-temperature annealed $x = 0.10$ sample is
very small from 300 K to 210 K in spite of a ferromagnetic
transition at about 300 K.  MR at 300 K is only of the order of 4\%
as compared to, 50\% observed by  S. H\'{e}bert et al \cite{SHeb}
and around 45\% to 66\% for increasingly cation deficient samples
studied by  Bukowaski et al \cite{ZBuk}. The low MR at 300K further
rules out the grain boundary effect as it would induce intergrain
tunneling MR. Moreover, the sudden increase in MR between 200 K and
80 K indicates that the ferromagnetic transition temperature lies
somewhere within this range. These facts clearly indicate that
ferromagnetic transition  at room temperature observed in
low-temperature annealed $x \ge 0.10$  samples is due to VRH
transport of charge carriers that connects regions of DE pairs
isolated from one another by non-magnetic Ti ions.

\subsection{XPS Studies}
\begin{figure}
\centering
\epsfig{file=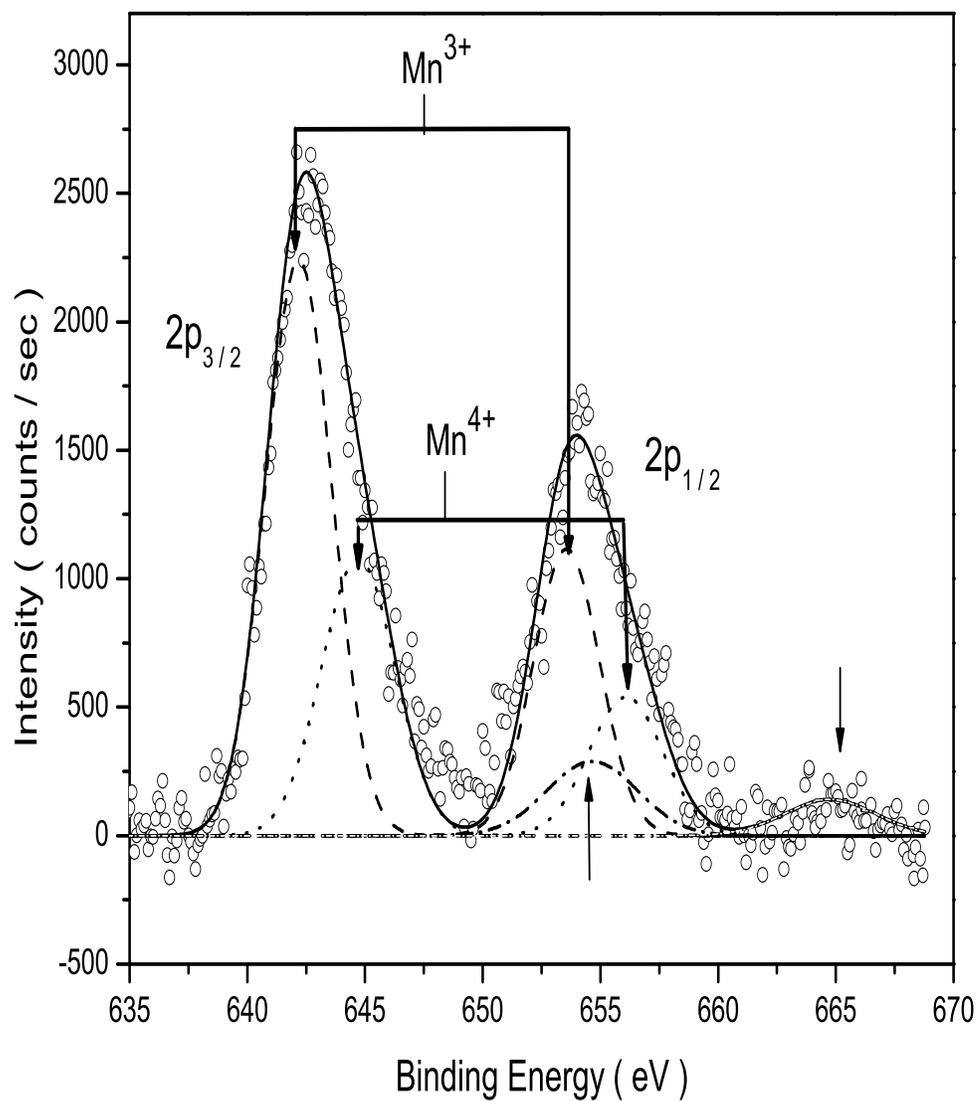, width=15cm, height=18cm}
\caption{XPS spectrum (circles) of La$_{0.67}$Sr$_{0.33}$MnO$_{3}$ sample along with fitted curve  (line). Discontinuous lines represent the deconvoluted 2p$_{3/2}$ and 2p$_{1/2}$ spectra corresponding to Mn$^{3+}$ and Mn$^{4+}$ ions. Arrows indicate the satellite peaks. }
 \label{fig:xps1}
\end{figure}

\begin{figure}
\centering
\epsfig{file=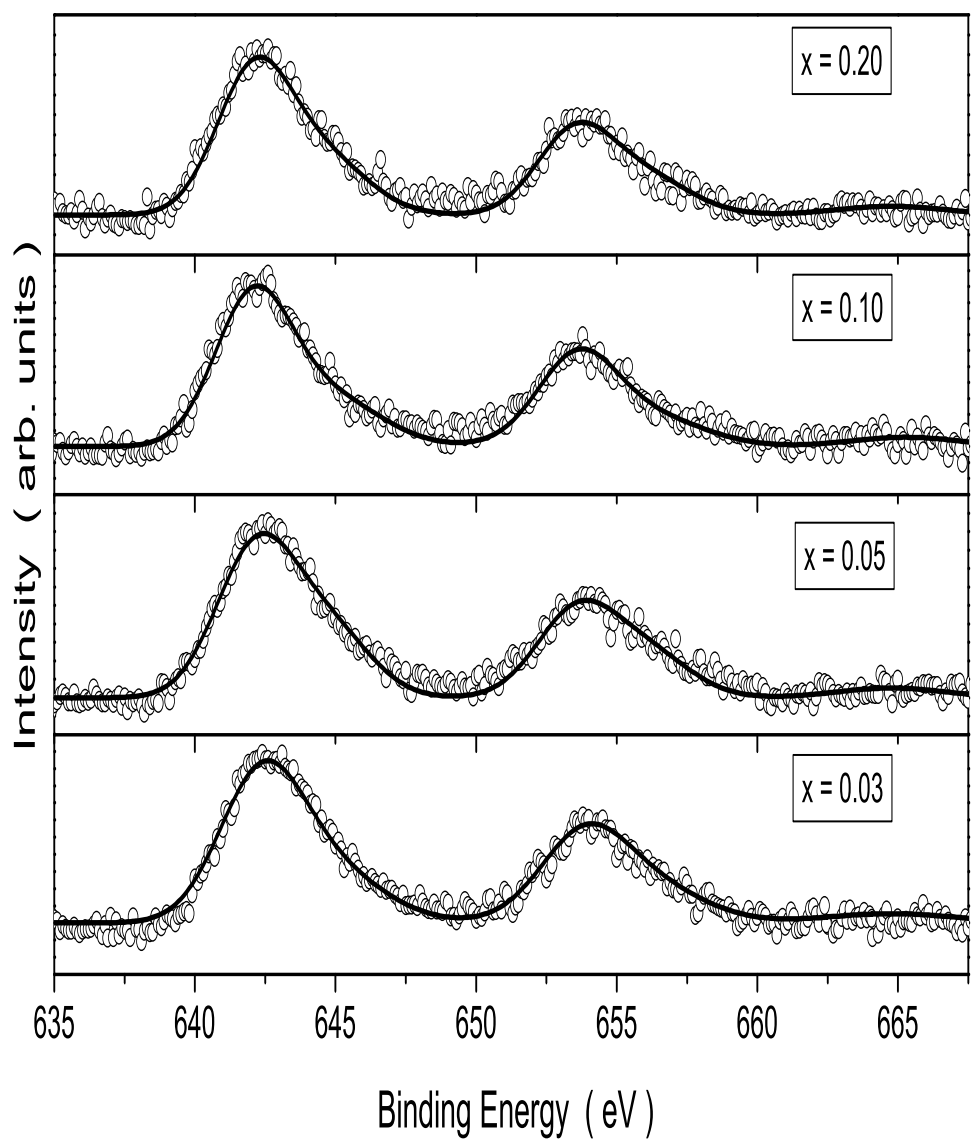, width=15cm, height=18cm}
\caption{XPS spectra (circles) for site-disordered La$_{0.67}$Sr$_{0.33}$Mn$_{1-x}$Ti$_x$O$_{3+\delta}$ samples along with the fitted curve (line)}
\label{fig:xps2}
\end{figure}

 From the above studies it is clear that in the low temperature
annealed samples, doping of Ti$^{4+}$ for Mn ions results in
isolated Mn rich regions wherein DE is active. If this is indeed the
case then the ratio of Mn$^{3+}$ to Mn$^{4+}$ should not increase
exponentially due to depletion of Mn$^{4+}$ by Ti$^{4+}$. To
estimate Mn$^{4+}$ and Mn$^{3+}$ contents, X-ray photoelectron
spectra of Mn 2p levels has been recorded for all the samples.
Figs.[\ref{fig:xps1} and \ref{fig:xps2}] show the Mn 2p X-ray
photoelectron spectra of the samples in the range $0.0 \le x \le
0.20$. The Mn 2p spectra exhibit two main peaks around 642.5 eV and
654  eV.  The two main peaks correspond to the spin-orbit split
2p$_{3\over2}$ and 2p$_{1\over2}$ levels, whereas the weak structure
at around 24 eV from the main peak is the satellite of the
2p$_{1\over2}$ peak.  The satellite of the 2p$_{3\over2}$ peak is
not visible because it overlaps with the 2p$_{1\over2}$ peaks.   The
two main peak features  were deconvoluted,   into those
corresponding to 2p$_{1\over2}$ and 2p$_{3\over2}$ states of
Mn$^{3+}$, Mn$^{4+}$ ions  and their satellites, with the help of a
curve fitting Peakfit software program. The peaks for
2p$_{3\over2}$ ions are higher in binding energy than those for
2p$_{1\over2}$ states  by about 11.6 eV. The  Mn$^{4+}$ and
Mn$^{3+}$ contents are calculated from areas under the curves
corresponding to Mn$^{4+}$ and Mn$^{3+}$ ions and are presented in
Table\ \ref{xps} . Results obtained clearly indicate the presence of
greater Mn$^{4+}$ content in the samples compared to a case wherein
Ti$^{4+}$ replaces only Mn$^{4+}$ and maintains charge balance.
Higher Mn$^{4+}$ content is possible only if substitution of
Ti$^{4+}$ is inhomogeneous such that it results in regions of sample
that are rich in Mn perhaps surrounded by Ti rich regions.

\begin{table}
 \caption{\label{xps} Percentage Mn$^{4+}$ and  Mn$^{3+}$ contents
in low-temperature annealed
La$_{0.67}$Sr$_{0.33}$Mn$_{1-x}$Ti$_x$O$_{3+\delta}$ samples
obtained from XPS data. Ti content was kept fixed to its nominal
value. Figures in the bracket indicate the maximum estimated
error.}
\begin{indented}
\item \begin{tabular}{@{}cccc}
  \br
Doping content ($x$) & Mn$^{4+}$ & Mn$^{3+}$ & Ti$^{4+}$ \\
\mr
0.03 & 32.2(0.9) & 64.8(0.9) & 3\\
0.05 & 32.9(0.7) & 64.1(0.7) & 5\\
0.10 & 27.7(0.8) & 62.3(0.8) & 10\\
0.20 & 23.7(0.8) & 56.3(0.8) & 20\\
\br
\end{tabular}
\end{indented}
\end{table}

\subsection{XANES}
\begin{figure}
 \centering
 \epsfig{file=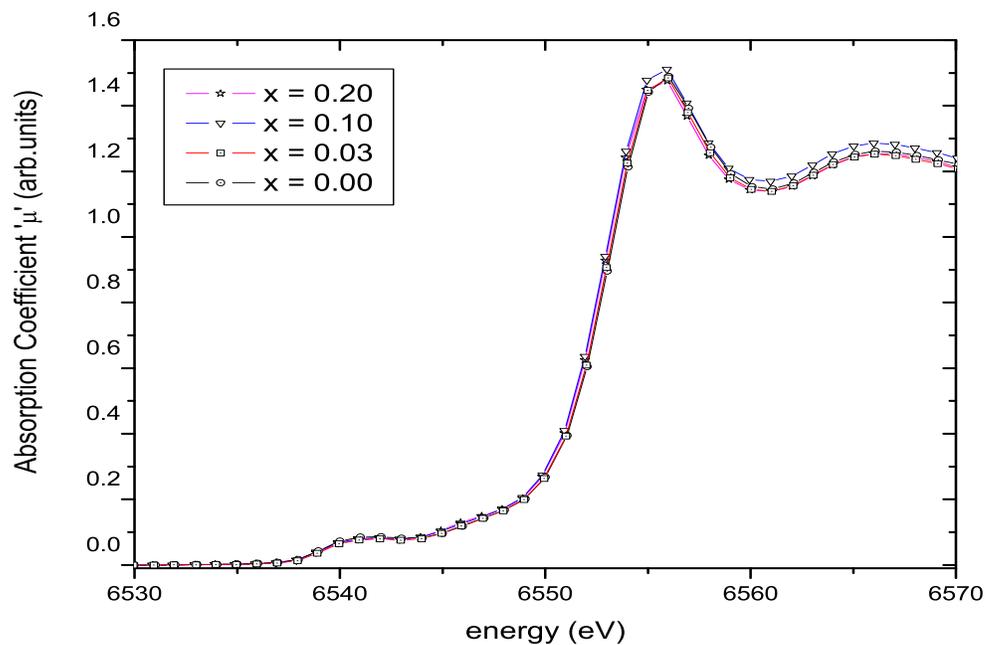, width=15cm, height=10cm}
  \caption{Normallised Xanes spectra of low-temperature annealed La$_{0.67}$Sr$_{0.33}$Mn$_{1-x}$Ti$_x$O$_{3+\delta}$ ($0 \le x \le 0.20$). }
  \label{fig:Mnxanes}
\end{figure}

 Normalized Mn K-edge XANES  spectra of
La$_{0.67}$Sr$_{0.33}$Mn$_{1-x}$Ti$_x$O$_{3+\delta}$ samples,
recorded at room temperature, are shown in Fig.\ \ref{fig:Mnxanes}.
The data was recorded with a step of 1eV, primarily for EXAFS
studies. The main peak arising due to 1s - 4p electronic transition
and a prepeak at about 15 eV below the main peak are clearly visible
in all the spectra. Positions of the main peak and the prepeak are
almost the same over the entire doping range $(0.00 \le x \le
0.20)$. All peak profiles are very similar to each other.
Surprisingly, chemical shift of the main edge is not observed over
the entire doping range. The chemical shift of the inflection point
of the main absorption edge is reported to be 4.2 eV for Mn$^{3+}$
and Mn$^{4+}$ ions in LaMnO$_{3}$ and CaMnO$_{3}$, respectively
\cite{GSUb,GDez}. If Ti$^{4+}$ ions substitute Mn$^{4+}$ ions alone,
the effective valence of the Mn ion will shift towards +3 with the
increasing doping content $x$ and such a change should be observed
in the form of shift of the main edge in Mn K-edge XANES spectra.
Such effect has been reported in Ti doped
La$_{0.7}$Ca$_{0.3}$Mn$_{1-x}$Ti$_x$O$_3$ samples \cite{DCao}. The
absence of  a chemical shift in our sample indicates that average
valence of Mn ions does not change appreciably in our samples which
again supports findings from XPS study.

\subsection{EXAFS}
EXAFS investigations of the low-temperature annealed  samples have
been undertaken to see the changes in octahedral environment of Mn
due to Ti substitution. Our interest lies mainly in La-(Mn/Ti), and
Mn-(Mn/Ti) bond lengths,  as these correlations can reveal whether
the substitution of Ti$^{4+}$ ions is homogeneous or not. Both La
and Mn K-edge EXAFS were fitted initially for
La$_{0.67}$Sr$_{0.33}$MnO$_3$ sample. Bond lengths  were varied
first, followed by  their Debye Waller Factors ($\sigma^2$).
Coordination numbers were initially fixed at the known
crystallographic values while fitting  Mn K-edge EXAFS and in case
of La K-edge EXAFS, twelve La-O bond lengths were divided into three
different groups of coordination numbers 3, 6 and 3 respectively on
the basis of closeness of their values. Later, coordination numbers
had to be varied slightly to improve the quality of fit. The Mn
K-edge and La K-edge EXAFS were fitted in tandem so that  equal
La-Mn bond lengths  were obtained from both the EXAFS.

\begin{figure}
\centering
\epsfig{file=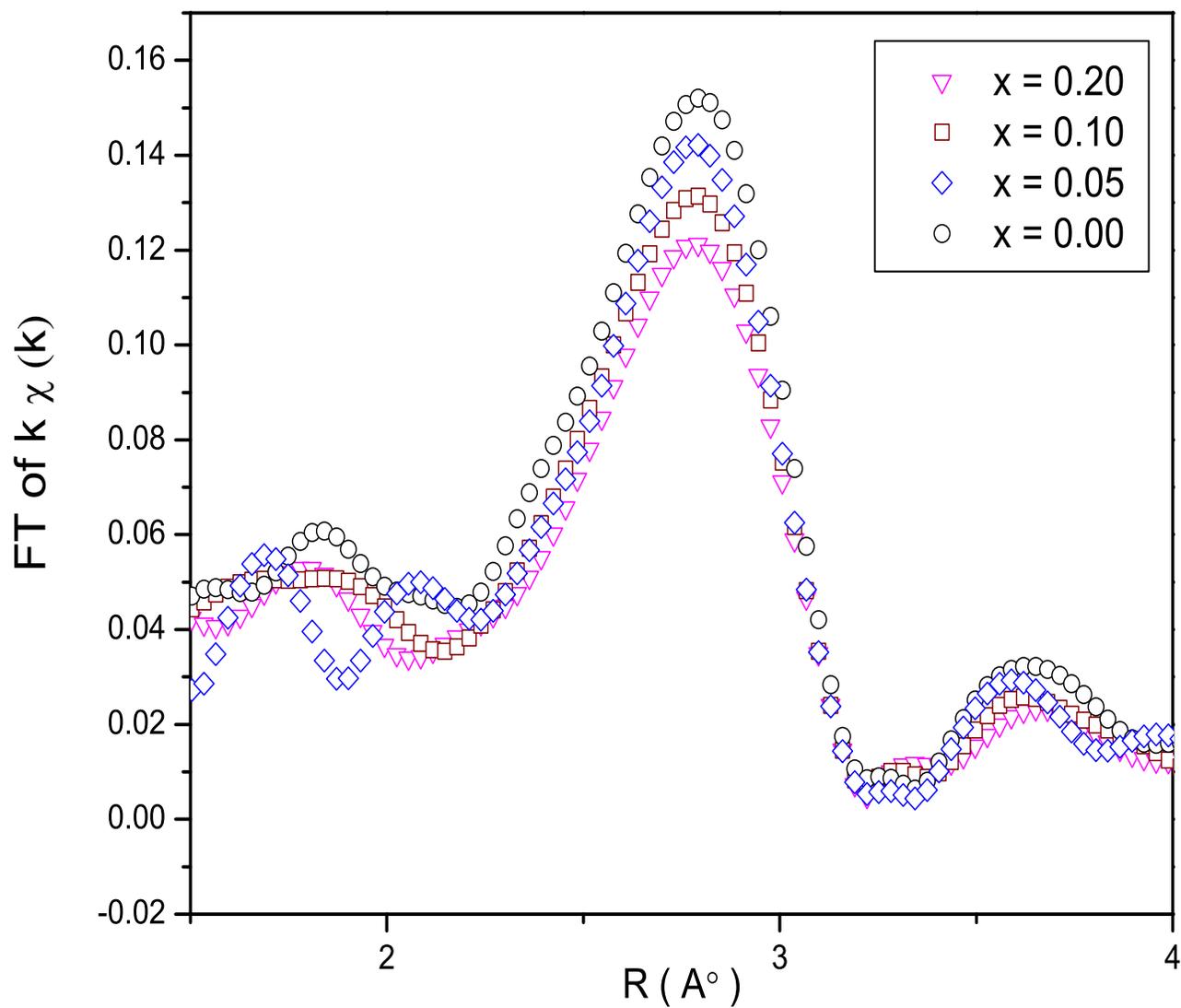, width=20cm, height=18cm}
\caption{La K-edge Fourier Transforms  of the low-temperature annealed La$_{0.67}$Sr$_{0.33}$Mn$_{1-x}$Ti$_x$O$_{3+\delta}$ samples. }
\label{fig:LaFT}
\end{figure}

\begin{figure}
\centering
 \epsfig{file=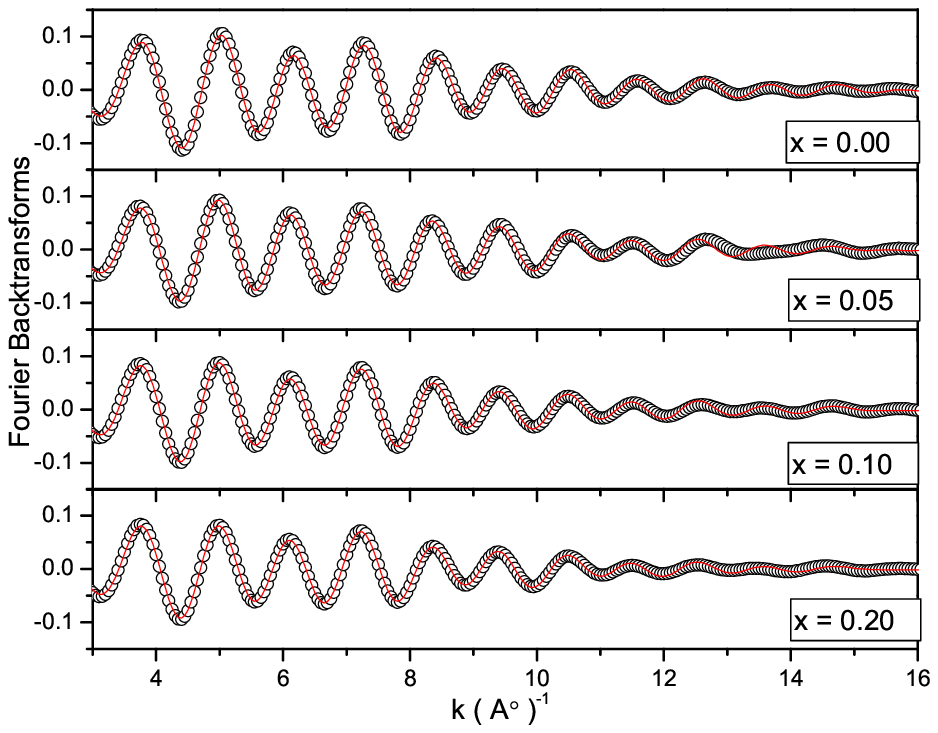, width=15cm, height=8cm}
  \caption{k - weighted backtransformed La K-edge EXAFS spectra (circles) along with fitted curves (line) in La$_{0.67}$Sr$_{0.33}$Mn$_{1-x}$Ti$_x$O$_{3+\delta}$.}
  \label{fig:Laqf}
\end{figure}

Fourier transforms of La K-edge EXAFS are presented in Fig.\
\ref{fig:LaFT}. The FT spectra are not corrected for phase shift,
however, the values of bond lengths reported in the text and tables
are phase corrected corrected values. La K-edge EXAFS were  fitted
with k-weighting = 1, in R-range 1.5 \AA~ to 3.5 \AA~ and k-range 3
\AA$^{-1}$ to 16 \AA$^{-1}$.   Fits of inverse Fourier transforms of
La K-edge EXAFS for  $x = 0.00, 0.05, 0.10$ and $0.20$ samples are
presented in Fig.\ \ref{fig:Laqf}. Results of La K-edge EXAFS
analysis are presented in Table\ \ref{LaKedge}. Nearly constant bond
lengths in these materials are in tune with the almost similar
lattice parameters obtained by analysis of X-ray diffraction spectra
of these materials [Table\ \ref{magn}].  As La-(Mn/Ti) correlation
in La K-edge EXAFS includes scattering from Mn$^{3+/4+}$ and
Ti$^{4+}$ ions, if Ti$^{4+}$ ions were to selectively replace
Mn$^{4+}$ ions, the bond length and $\sigma^2$ of this correlation
should have systematically increased with doping content $x$.
However, no such change is seen either in the bond length or
$\sigma^2$.Furthermore, the changes in La-O bond lengths and their
Debye Waller Factors are small and non-systematic and  can be
understood in terms of random distribution of Ti$^{4+}$, Mn$^{3+}$
and Mn$^{4+}$ ions in the sample and resulting oxygen displacement
that is permissible in rhombohedral (R\={3}C) structure. In this
structure, effect of  oxygen displacements will be evident, more on
La-O$_1$ and La$O_3$ bonds compared to that on La-O$_2$ which can be
seen from the values reported in Table \ref{LaKedge}.

\begin{figure}
 \centering
 \epsfig{file=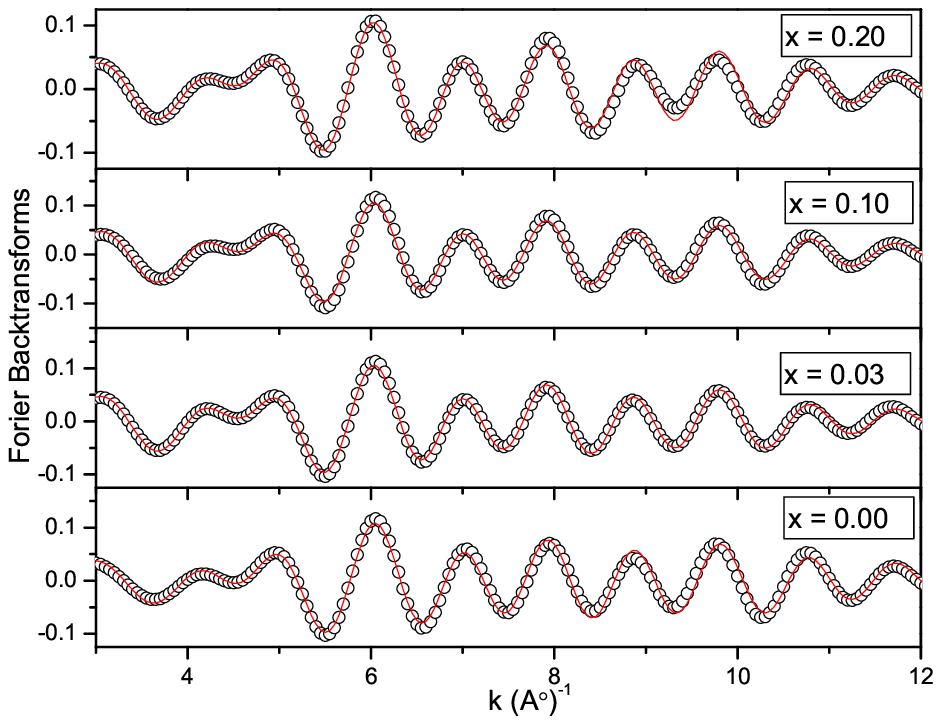, width=15cm, height=10cm}
  \caption{k - weighted backtransformed Mn K-edge EXAFS spectra (circles) along with fitted curves (line) in La$_{0.67}$Sr$_{0.33}$Mn$_{1-x}$Ti$_x$O$_{3+\delta}$. }
  \label{fig:Mnqf}
\end{figure}

Mn K-edge EXAFS were fitted with k-weighting, in R-range 1\AA~ to
3.6\AA~ and k-range 3\AA~ to 12\AA$^{-1}$.  Fits of the inverse
Fourier transforms of Mn K-edge EXAFS for  $x$ = 0.00, 0.03, 0.10
and 0.20 are presented in Fig.\ \ref{fig:Mnqf} and the results of
the Mn K-edge analysis are presented in Table\ \ref{MnKedge}. Mn-O,
Mn-Mn as well as Mn-La bond lengths for the samples, obtained by
fitting Mn K-edge EXAFS, are almost the same  over the entire doping
range $0.00 \le x \le 0.20$.  Ti substitution should have seriously
affected the Mn-O bond length at least at 10\% and 20\% doping level
due to ionic size difference. Therefore a constant bond length could
only imply that for most Mn ions, the local environment remains
unaltered. This is possible only if in the low temperature annealed
samples the Ti substitution is inhomogeneous leading to creation of
Mn rich regions over bulk of the sample separated by Ti rich
regions. This is further supported from the fact that  Mn-(Mn/Ti)
distance obtained from Mn K-edge EXAFS and La-Mn distance obtained
from Mn K-edge as well as La K-edge EXAFS remains almost the same
throughout the series. Therefore in the case of inhomogeneous
doping, even at high doping level of 20\%, the local environment
around La and Mn ions would be similar to that in the undoped sample
giving constant bond lengths through out the series.

\begin{table}
 \caption {\label{LaKedge} Structural parameters obtained from La
K-edge EXAFS  for low-temperature annealed
La$_{0.67}$Sr$_{0.33}$Mn$_{1-x}$Ti$_x$O$_{3+\delta}$ samples.
Figures in brackets indicate uncertainties  in the last digit. }
\begin{indented}
\item \begin{tabular}{@{}cccccc}
  \br
&  & x=0.00 & x=0.05& x=0.10& x=0.20 \\
\mr
La-O$_1$ &R(\AA)&2.48(1)&2.43(1)&2.45(1)&2.44(1)\\
& $\sigma^2$(\AA$^2$)  &0.008(2)&0.007(2)&0.006(1)&0.005(1)\\
&n&3.3(5)&3.3(5)&3.3(3)&3.3(2)\\
La-O$_2$ &R(\AA)&2.64(1)&2.63(1)&2.63(1)&~~2.62(1)\\
& $\sigma^2$(\AA$^2$)  &0.010(2)&0.008(3)&0.008(1)&0.009(1)\\
&n&6.2(7)&6.2(6)&6.2(4)&6.2(3)\\
La-O$_3$ &R(\AA)&2.84(2)&2.85(3)&2.86(1)&2.84(1)\\
& $\sigma^2$(\AA$^2$) &0.005(2)&0.006(3)&0.006(6)&0.009(2)\\
&n&2.7(6)&2.7(7)&2.7(3)&2.7(5)\\
La-Mn/Ti &R(\AA)&3.39(1)&3.39(1)&3.39(1)&3.39(1)\\
& $\sigma^2$(\AA$^2$)  &0.005(1)&0.005(1)&0.006(1)&0.007(1)\\
&n&8.1(4)&8.1(4)&8.1(2)&8.1(2)\\
\br
\end{tabular}
\end{indented}
\end{table}

\begin{table}
 \caption {Structural parameters obtained from Mn K-edge EXAFS for
low-temperature annealed
La$_{0.67}$Sr$_{0.33}$Mn$_{1-x}$Ti$_x$O$_{3+\delta}$ samples.
Figures in brackets indicate uncertainties  in the last
digit.\label{MnKedge}}
\begin{indented}
\item \begin{tabular}{@{}cccccc}
\br
&& x=0.00 &x=0.03 &x=0.10 & x=0.20 \\
\mr
Mn-O &{\footnotesize R(\AA)}&1.92(0)&1.92(0)&1.92(1)&1.92(1)\\
& {\footnotesize $\sigma^2$(\AA$^2$) } &{\footnotesize 0.007(2)}&{\footnotesize 0.007(1)}&{\footnotesize 0.007(2)}&{\footnotesize 0.009(1)}\\
&n&5.5(6)&5.5(3)&5.5(4)&5.7(5)\\
Mn-(La/Sr)&{\footnotesize R(\AA)}&3.39(1)&3.39(0)&3.39(0)&3.39(0)\\
& {\footnotesize $\sigma^2$(\AA$^2$)}  &{\footnotesize 0.009(1)}&{\footnotesize 0.010(1)}&{\footnotesize 0.010(1)}&{\footnotesize 0.010(1)}\\
&n&8.4(7)&8.4(5)&8.4(5)&8.6(5)\\
Mn-(Mn/Ti) &{\footnotesize R(\AA)}&3.81(1)&3.82(0)&3.81(0)&3.82(0)\\
& {\footnotesize $\sigma^2$}  &{\footnotesize 0.004(1)}&{\footnotesize 0.002(0)}&{\footnotesize 0.003(1)}&{\footnotesize 0.002(1)}\\
&n&5.6(5)&5.6(2)&5.6(2)&5.6(2)\\
Mn-O-(Mn/Ti) &{\footnotesize R(\AA)}&3.83(1)&3.83(1)&3.83(1)&3.83(1)\\
& {\footnotesize $\sigma^2$(\AA$^2$) } &{\footnotesize 0.007(5)}&{\footnotesize 0.007()}&{\footnotesize 0.007(1)}&{\footnotesize 0.007(0)}\\
&n&12.4(6)&12.4(3)&12.4(3)&12.3(4)\\
\br
\end{tabular}
\end{indented}
\end{table}

\section{Discussion}
 Structural, transport and magnetic properties of Ti doped
low-temperature annealed  La$_{0.67}$Sr$_{0.33}$MnO$_{3}$ samples
have been investigated in the present study. Samples with 10 \% Ti
doping were prepared using two different preparation  schedules.
High temperature annealed  sample shows transport and magnetic
properties similar to those reported in \cite{kal} whereas low
temperature annealed sample has significantly different properties.
With increasing doping content $x$, substitution of Ti$^{4+}$
(3d$^{0}$) ions in Mn$^{4+}$-O-Mn$^{3+}$ chains is expected to
increase resistivity of the samples. However, low-temperature and
high-temperature annealed $x = 0.10$ samples exhibit significantly
different resistivities. Moreover, high-temperature annealed $x =
0.10$ sample exhibits metal-insulator transition whereas
low-temperature annealed sample is insulating over the entire
temperature range investigated.  This fact can be understood in
terms of additional scattering centres like cation vacancies or
substitutional inhomogeneities in the low-temperature annealed
sample.  The short-range order of Mn$^{4+}$-O-Mn$^{3+}$ chain,
caused by interspersed Ti ions and cation vacancies, gives rise to
the insulating behavior of low-temperature annealed  x $\ge$ 0.10
samples. Strong localizing effect of random cation vacancies is well
known \cite{Coey}. The random potential fluctuations due to cation
vacancies and dopant Ti ions favor Anderson type localization, but
the localized wavepackets are large enough for the e$_g$ electrons
to extend over several sites to provide the ferromagnetic
interaction. The samples with x $\le$ 0.05  exhibit  sharp rise in
susceptibility  indicating paramagnetic to  ferromagnetic
transition and  the ferromagnetic ordering temperature  ($T_C$)
decreases with increasing doping content, indicating the  gradual
weakening of double-exchange coupling.  However, for $ x = 0.10$ and
$0.20$  samples, T$_C$ remains at  around 300 K, which is
significantly higher than those reported for  Ti doped manganites
\cite{Xian,sah,kal,kim,hu}.  Higher T$_C$ of these samples can not
be simply attributed to the presence of cation vacancies. It is to
be noted that  T$_C$ decreases with increasing cation deficiency in
La$_{1-x}$Mn$_{1-y}$O$_3$ \cite{PSIP}. In case of
La$_{0.815}$Sr$_{0.185}$MnO$_{3+\delta}$ samples, T$_C$ shifts
marginally on the higher temperature side for small $\delta$ and
shifts towards lower temperatures for higher $\delta$ \cite{ZBuk}.
Hence,  high T$_C$ in our low temperature annealed $x = 0.10$ and $x
= 0.20$   samples may not be due to cation vacancies alone, it could
be   related to Ti doping as well. Moreover, susceptibility
(emu/f.u.) of these   $x = 0.10$ and $0.20$  samples is
significantly smaller  and  can not be explained only on the basis
of  substitution of Mn$^{4+}$ ions by non-magnetic Ti$^{4+}$(3d$^0$)
ions. Decrease in susceptibility of such a magnitude  is not
observed in Ti doped samples of the same composition
\cite{kal,kim,troy}. T$_C$ of our high-temperature annealed  $x =
0.10$ sample matches closely with the reported value of T$_C$
\cite{kal}. In fact susceptibility is found to be higher in cation
deficient samples than that in stoichiometric samples\cite{LRan} and
similar magnitudes in\cite{ZBuk}. Reduced susceptibility in our
low-temperature annealed samples can be explained to be due to
formation of isolated Mn rich regions where in DE is active
separated by Ti rich regions.These regions are created due to
inhomogeneous Ti substitution.

IR spectra indicate that Ti is predominantly in tetravalent state in
these compounds. Moreover, IR spectra exhibit a slight shift in the
stretching mode absorption peak in our samples. It has been argued
on the basis of fixed Mn-O stretching  mode absorption peaks in
La$_{0.67}$Ca$_{0.33}$Mn$_{1-x}$Ti$_{x}$O$_{3}$ that Jahn-Teller
distortions are not affected by doping of Ti ions in their
materials\cite{Xian}. In the case of low temperature annealed
samples, the 600cm$^{-1}$ mode is a quite broadened. This can be
related to the presence of disorder in the Ti doping. In this sample
Ti$^{4+}$ substitution causes regions  wherein double exchange is
active and regions which are devoid of double exchange pairs.
Perhaps the two regions are in such a proportion that the strengths
of respective stretching modes are nearly same resulting in a broad
absorption dip. XRD studies indicate the  complete solid solubility
of Ti in the parent R\={3}C structure of
La$_{0.67}$Sr$_{0.33}$MnO$_{3}$ with lattice parameters of the
low-temperature annealed samples remaining  nearly the same over the
entire doping range. This also supports the above idea of
inhomogeneous substitution of Ti$^{4+}$ ions. Such a substitution
would result in formation of Mn rich regions that are isolated from
each other by regions that are rich in Ti. This would increase the
magnitude of resistivity due to additional scattering from the Ti
rich regions, inhibit insulator to metal transition, suppress
magnetic order as well as magnetoresistance,  all of which are seen
in low temperature annealed $x$ = 0.1 and 0.2 samples. The weak
ferromagnetic transition seen in these two samples at about 300K
could then be explained to be due to ordering of Mn rich regions
connected with each other by a variable range hopping polaron. The
VRH of charge carriers strengthens ferromagnetic exchange-coupling
between isolated ferromagnetic regions and  manifests in a weak
ferromagnetic-like transition at room temperature.

EXAFS results provide the further evidence for the inhomogeneous
substitution of Ti ions. Local structure around La and Mn ions is
found to be almost the same in doped as well as the undoped samples
[Table (\ref{LaKedge}), (\ref{MnKedge})]. This indicates that even
at 20\% dopant concentration, large parts of the sample have Mn ions
in exactly similar environment as in the undoped sample. In
low-temperature annealed $x = 0.10$  sample,  Mn$^{4+}$ and
Mn$^{3+}$ contents  are found  to be nearly 28\% and 62\%
respectively [Table\ \ref{xps}]. If the substitution was
homogeneous, 10\% Ti ions would have depleted the Mn$^{4+}$
concentration so as to maintain charge balance. Instead, in the low
temperature annealed samples, it can be seen that Mn$^{4+}$
concentration decreases at much slower rate indicating the presence
of isolated Mn rich regions.

\section{Conclusions}

In this paper studies of structural, magnetic, transport and
spectroscopic properties of low temperature annealed
La$_{0.67}$Sr$_{0.33}$Mn$_{1-x}$Ti$_x$O$_{3}$  ($0 \le x \le 0.20$)
is presented. The lower annealing temperature results in an
inhomogeneous substitution of Ti ions resulting in a nearly similar
crystal structure and local structure around La and Mn ions but
vastly different magnetic and transport properties as compared to
the undoped sample. This is perhaps due to the formation of isolated
ferromagnetic clusters linked to each other by a variable range
hopping polaron.

\ack
K R Priolkar acknowledges financial assistance from DST under SR/FTP/PS-19/2003.
Thanks are also due, to Prof M S Hegde, Indian Institute of Science,
Bangalore for useful discussions and constant encouragement and to
Dr N Y Vasanthacharya for DC susceptibility measurements. XAFS
measurements were performed at SPring-8 under the proposal no. {\bf
2003A0028-Cx-np} with financial assistance for travel from DST, New
Delhi.

\newpage

\end{document}